# Contribution of defects to the spin relaxation in copper nanowires

Estitxu Villamor,[1] Miren Isasa,[1] Luis E. Hueso,[1,2] and Fèlix Casanova[1,2]

[1] CIC nanoGUNE, 20018 Donostia-San Sebastian, Basque Country (Spain)
[2] IKERBASQUE, Basque Foundation for Science, 48011 Bilbao, Basque Country (Spain)

The contributions to the spin relaxation in copper (Cu) nanowires are quantified by carefully analyzing measurements of both charge and spin transport in lateral spin valves as a function of temperature and thickness. The temperature dependence of the spin-flip scattering solely arises from the scattering with phonons, as in bulk Cu, whereas we identify grain boundaries as the main temperature-independent contribution of the defects in the nanowires. A puzzling maximum in the spin diffusion length of Cu at low temperatures is found, which can be explained by the presence of magnetic impurities. The results presented here suggest routes for improving spin transport in metallic nanostructures, otherwise limited by confinement effects.

## I. INTRODUCTION

In the last two decades, *spintronics* has attracted a great deal of attention because of its potential application to information technology. The advantages of this emerging field, which is based on the transport of the quantum spin of the electron, include faster data processing speed, non-volatility and lower electrical power consumption as compared to conventional electronics [1,2]. Devices which are capable of manipulating and transporting spins for long distances are crucial [3]. Therefore, it is of uttermost importance to understand the spin relaxation processes leading to the loss of the spin information inside a non-magnetic material, including metals [4-6], semiconductors [7-9], graphene [10,11] or organic materials [12]. An attractive way to investigate such processes is by the use of pure spin currents, which allow studying the spin transport mechanisms without the presence of spurious effects caused by the charge currents [13].

Whereas nanostructures are needed in order to be able to electrically generate and detect pure spin currents, the inherent confinement related to such nanostructures introduces additional sources of spin relaxation. In a non-magnetic material, spin relaxation arises both from the scattering with phonons and the defects of the material, which include impurities, grain boundaries and the surface. In particular, the role of the surface to the spin-flip scattering has recently been debated due to its effect to the spin diffusion length at low temperatures [5,6,14,15]. In this letter, by analyzing the charge and spin transport in copper (Cu) nanowires as a function of the temperature and thickness, we quantify the relative importance of each contribution to the spin relaxation through the corresponding spin-flip probabilities. This work will help us to identify ways to improve spin transport in metallic nanodevices.

With this purpose, we have used lateral spin valves (LSV), devices which allow us to electrically inject and detect a pure spin current in a non-magnetic (NM) nanowire by using ferromagnetic (FM) electrodes in a non-local configuration [4-6,13-23].

## II. EXPERIMENTAL DETAILS

LSV devices were fabricated on SiO$_2$ (150 nm)/Si substrates by a two-step *e*-beam lithography, UHV evaporation and lift-off process. In the first step, FM electrodes were patterned in PMMA resist, and Ni$_{80}$Fe$_{20}$ (permalloy, Py) was *e*-beam evaporated with a base pressure ≤1·10$^{-8}$ mbar. In the second lithography step, the NM nanowire and contact pads were patterned, and Cu was thermally evaporated on top of the Py electrodes with the same base pressure ≤1·10$^{-8}$ mbar. Ar-ion milling was performed prior to the Cu deposition in order to remove resist left-overs and any oxide layer from the Py electrodes, with optimized conditions to obtain a clean, transparent interface. Five samples were fabricated, with Cu thicknesses (*th*) of 145, 100, 70, 60 and 40 nm, and resistivities ($\rho_{Cu}$) of 1.18, 1.26, 1.63, 1.98 and 2.22 μΩ cm at 10 K, respectively (see Table 1). The width of the Cu nanowire is ~200 nm for all samples. The thickness of Py is 35 nm for all samples and the resistivity, which was measured in a different device deposited under the same nominal conditions, is 22.4 μΩ cm (32.9 μΩ cm) at 10 K (300K). Different widths of the Py electrodes, ~110 and ~150 nm, were chosen in order to obtain different switching magnetic fields. Each sample contains up to 10 different LSVs with an edge-to-edge separation distance (*L*) between Py electrodes from 200 to 3500 nm. Figure 1(a) is a scanning electron microscopy (SEM) partial image of a sample, showing two devices.

Non-local spin valve measurements were performed using a "dc reversal" technique [16] in a liquid He cryostat, with applied magnetic field *H* and temperatures ranging from 10 to 300 K. The measured voltage *V*, normalized to the absolute value of the current *I*, is defined as the non-local resistance $R_{NL} = V/|I|$ (see Fig. 1(a) for a measurement scheme). This value changes from positive to negative when the relative magnetization of the FM electrodes changes from parallel to antiparallel by sweeping *H*. The difference between the positive and negative values of $R_{NL}$ is defined as the spin signal, $\Delta R_{NL}$, which is proportional to the spin accumulation at the detector (see Fig. 1(b)).

An expression for the spin signal is obtained by applying the one-dimensional spin-diffusion model to our geometry [16,18,24]:

$$\Delta R_{NL} = \frac{2\alpha_{Py}^2 R_{Cu}}{\left(2+\frac{R_{Cu}}{R_{Py}}\right)^2 e^{L/\lambda_{Cu}} - \left(\frac{R_{Cu}}{R_{Py}}\right)^2 e^{-L/\lambda_{Cu}}}, \quad (1)$$

where $\alpha_{Py}$ is the spin polarization of the Py, $R_{Cu} = 2\lambda_{Cu}\rho_{Cu}/S_{Cu}$ and $R_{Py} = 2\lambda_{Py}\rho_{Py}/S_{Py}(1-\alpha_{Py}^2)$ are the spin resistances, $\lambda_{Cu,Py}$ the spin diffusion lengths, $\rho_{Cu,Py}$ the resistivities and $S_{Cu,Py}$ the cross-sectional areas of Cu and Py, respectively. For all the devices, we use $\lambda_{Py}$ = 5 nm at 10 K [14, 25] and consider a temperature dependence coming from the resistivity in the form $\lambda_{Py} = const./\rho_{Py}$. This approximation is satisfactory up to ~100 K. Above that temperature, the contribution from magnons lowers $\lambda_{Py}$ and the values obtained with the previous formula are an upper estimate [4,25]. At 300 K we estimate $\lambda_{Py}$ = 3.4 nm, which is nonetheless in good agreement with the values reported in previous works [14, 18, 25]. Geometrical parameters were measured by Scanning Electron Microscopy (SEM) for each device. $\Delta R_{NL}$ as a function of the distance $L$ is fitted to Eq. 1 for each sample, with $\alpha_{Py}$ and $\lambda_{Cu}$ as fitting parameters (see Fig. 1(c)). Due to the wide range of distances $L$ that we use, the two values are completely uncoupled. For all samples, we obtain the values of $\alpha_{Py}$, which are consistently between 0.38 and 0.40 (0.31 and 0.34) at 10 K (300 K), in agreement with literature [6,13,14,16-18].

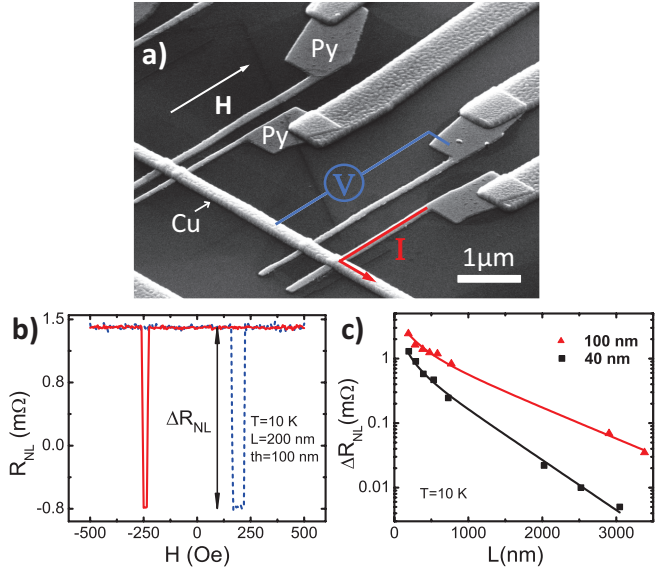

Figure 1: (a) SEM image of a typical sample, showing two lateral spin valves. The non-local measurement configuration, materials (Py and Cu) and the direction of the applied magnetic field $H$ are shown. (b) Non-local resistance, measured at 10 K, for a LSV with $L$=200 nm and $th$=100 nm. Solid red (dashed blue) line indicates the decreasing (increasing) direction of the magnetic field. Spin signal is tagged as $\Delta R_{NL}$. (c) Spin signal as a function of the distance $L$ between Py electrodes measured at 10 K for two samples with $th$=100 nm (red triangles) and 40 nm (black squares). Solid lines are fits to Eq. 1.

## III. RESULTS AND DISCUSSION

### A. Charge transport in Cu nanowires

For a proper analysis of $\lambda_{Cu}$ values, we first need to characterize the resistivity of the Cu nanowire. It was measured as a function of temperature for each sample (only three curves are shown in Fig. 2 for the sake of clarity), showing all curves the same temperature dependence. The phonon-scattering contribution to the resistivity is described by the Bloch-Grüneisen theory:

$$\rho(T) = \rho_0 + K\left(\frac{T}{\Theta}\right)^5 \int_0^{\Theta/T} \frac{x^5}{(e^x-1)(1-e^{-x})} dx , \qquad (2)$$

where $\rho_0$ is the residual resistivity, $K$ is a constant for a given metal and $\Theta$ is the Debye temperature [26,27]. We obtain $\Theta$ ~280 K and $K$~6.5 µΩ cm for all samples when fitting the experimental curves to Eq. 2, in agreement with previously reported values [27,28]. The residual resistivity, arising from scattering with defects, is thus temperature independent and increases with decreasing the Cu thickness as can be seen in Fig. 2. Both grain-boundary scattering and surface scattering are known to increase the resistivity with decreasing the thickness of thin films and nanowires. However, whereas the former contribution does not change the temperature dependence of the resistivity with respect to the bulk one, the latter induces a deviation in this temperature dependence [29-31]. This indicates that the defect scattering of the Cu nanowires is dominated by grain boundaries rather than surface.

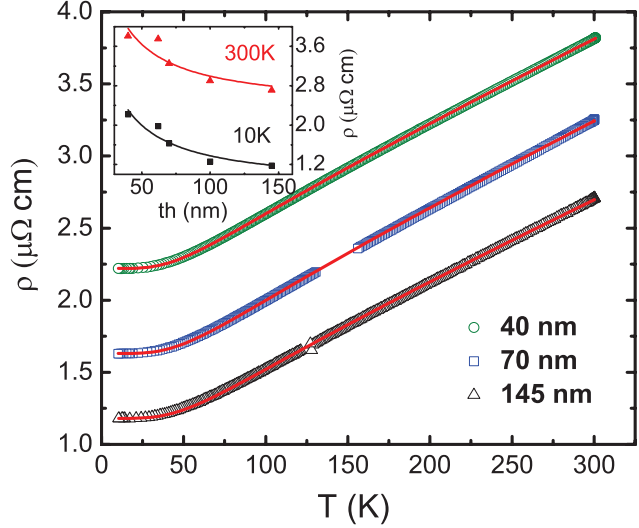

Figure 2: Resistivity as a function of temperature for Cu nanowires with different thicknesses. Red solid lines are fits to the Bloch-Grüneisen equation. Inset: Resistivity of the Cu channels as a function of the thickness at 10 K (black squares) and 300 K (red triangles). Solid lines are fits to the Mayadas and Shatzkes model.

The grain-boundary-dominated scattering is described by the Mayadas and Shatzkes model [29-33]. This model, to a first approximation, has a dependence with the inverse of the average grain size $d$ by $\rho = \rho_b \left(1 + \frac{3\ell}{2}\frac{R}{1-R}\frac{1}{d}\right)$, where $\rho_b$ and $\ell$ are the resistivity and the mean free path of the bulk Cu, and $R$ is the reflection coefficient of the electrons colliding at the grain boundaries. One can consider that, in evaporated Cu, $d$ is given by the smallest dimension of the wire, *i.e.* the thickness in this case [29], and fit the resistivity of Cu as a function of the thickness to the Mayadas and Shatzkes model, as it is shown in the inset of Fig. 2. This fitting is consistent both at 10 K and 300 K, yielding a temperature independent value of $R$ = 0.38, which is in good agreement with literature [29,31,33]. From the fitting, we also obtain the value of $\rho_b$= 0.73 μΩ cm (2.31 μΩ cm) at 10 K (300 K), slightly larger than pure bulk Cu resistivity, due to the likely presence of other impurities [33].

**B. Spin transport in Cu nanowires**

Figure 3(a) shows the values of the spin diffusion length of Cu, obtained from the fitting of Eq. 1, as a function of the temperature. The highest values of the spin diffusion length correspond to the samples with the thickest Cu nanowire (145 nm). A $\lambda_{Cu}$ of 1020 ± 6 nm is obtained at 10 K, being in good agreement with the largest values reported [4,13,14]. For the sample with a 40-nm-thick Cu nanowire, a $\lambda_{Cu}$ of 500 ± 16 nm is obtained at 10 K. The values of $\lambda_{Cu}$ tend to increase with thickness, following the opposite trend of the resistivity (see Table 1). In fact, $\lambda_{Cu}$ shows, with a slight dispersion, a linear dependence with the inverse of the resistivity at 10 K, (see Fig. 3(b)). Figure 3(a) also shows a maximum in $\lambda_{Cu}$ around 30 K for all thicknesses. This behavior, which cannot be explained by the Elliott-Yafet mechanism for spin relaxation [34,35], has been previously reported in Cu [14-16] and in silver (Ag) LSV devices [5,6], and is discussed below.

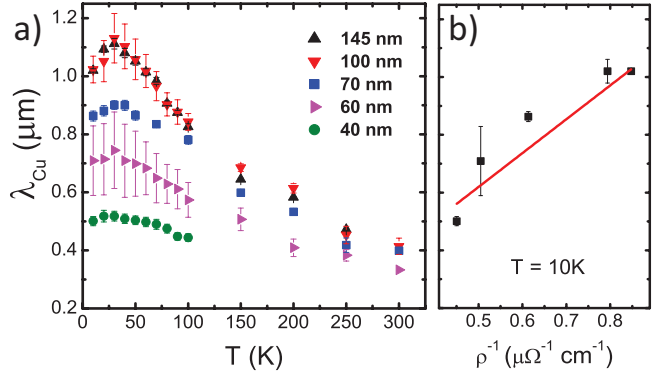

Figure 3: (a) Spin diffusion length of Cu as a function of temperature obtained for nanowires with different thicknesses. (b) Spin diffusion length of Cu as a function of the inverse of the resistivity, at 10 K. Red solid line is a linear fit of the data.

The spin diffusion length is represented by $\sqrt{D\tau_{sf}}$, where $D = 1/N(E_F)e^2\rho$ is the diffusion constant, $N(E_F)$ is the electronic density of states at the Fermi level ($1.8\times10^{28}$ states/eV/m$^3$ for Cu [4]), $e$ is the electronic charge and $\tau_{sf}$ the spin relaxation time of the NM metal. The spin relaxation mechanism in metals arises from the spin-orbit interaction, as discussed by Elliott and Yafet [34,35]. In this case, the spin relaxation time is proportional to the momentum relaxation time $\tau_e$ by the spin-flip probability $a$ in the form $1/\tau_{sf} = a/\tau_e$.

The momentum relaxation time can be calculated from the measured resistivity with $\tau_e = 3/(v_F^2 N(E_F)e^2\rho)$, where $v_F$ is the Fermi velocity ($1.57\times10^6$ m/s for Cu [4]), and decomposed in two different contributions coming from the phonons *ph* and the defects *def* (including impurities, grain boundaries and surface) using the Matthiessen rule $(\tau_e)^{-1} = \left(\tau_e^{ph}\right)^{-1} + \left(\tau_e^{def}\right)^{-1}$. As discussed previously in the analysis of the resistivity, the first contribution is temperature-dependent whereas the second one is temperature-independent. Accordingly, the spin relaxation rate $(\tau_{sf})^{-1}$ can be expressed as

$$\frac{1}{\tau_{sf}} = \frac{a_{ph}}{\tau_e^{ph}} + \frac{a_{def}}{\tau_e^{def}}, \qquad (3)$$

where $a_{ph}$ and $a_{def}$ are the spin-flip probabilities for each contribution.

Figure 4 shows, for all samples, $(\tau_{sf})^{-1}$ as a function of the phonon scattering rate, $(\tau_e^{ph})^{-1}$, which has been calculated from the $\rho(T)$ measurements (Fig.

2), the Matthiessen rule and taking into account that the residual resistivity gives the defect scattering rate, $(\tau_e^{def})^{-1}$ (listed in Table 1).

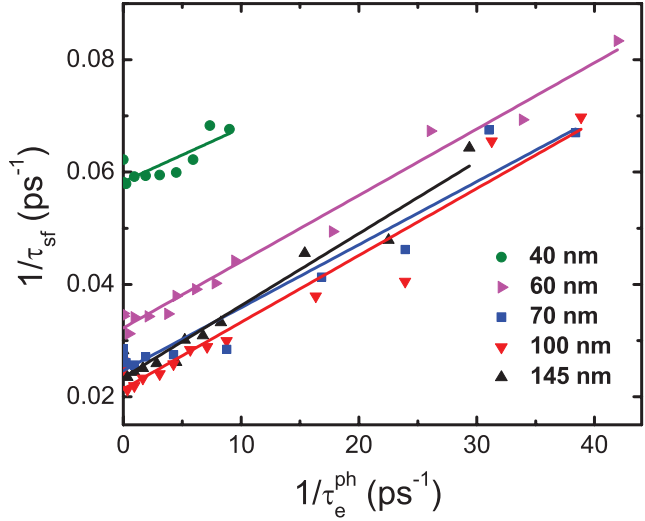

Figure 4: Spin relaxation rate as a function of the phonon scattering rate. Symbols are experimental data and solid lines are linear fits to Eq. 3.

The minimum in $(\tau_{sf})^{-1}$ associated to the maximum in $\lambda_{Cu}$ (Fig. 3), is smeared out in this representation, and a clear linear dependence of the experimental data is observed for all thicknesses, which can be fitted to Eq. 3. The value of $a_{ph}$ is directly obtained from the slope. The intercept to the Y axis corresponds to $a_{def}/\tau_e^{def}$, which is the contribution of the defects to the total spin relaxation rate $(\tau_{sf})^{-1}$. Since we already know the value of $(\tau_e^{def})^{-1}$, we can independently obtain $a_{def}$.

The values of the spin-flip probabilities $a_{ph}$ and $a_{def}$ for all samples are shown in Table 1. Cu nanowires with different thicknesses consistently yield the same value of $a_{ph}$ (~1.1×10$^{-3}$), which is an intrinsic parameter of bulk Cu. Similar values have been previously obtained in bulk Cu using conduction-electron spin resonance (CESR) experiments ($a_{ph}$ = 1.1×10$^{-3}$) [36], as well as in Cu nanowires with LSV experiments ($a_{ph}$ = 2.0×10$^{-3}$) [4]. The value of $a_{def}$ is very similar for all thicknesses as well, evidencing that the nature of the defects contributing to the spin relaxation is the same in all samples. This is consistent with the observed linear dependence of $\lambda_{Cu}$ with the inverse of resistivity at 10 K (Fig. 3(b)). These defects are mostly grain boundaries, as demonstrated from the resistivity analysis, although the slight dispersion in the values of $a_{def}$ also suggests small differences in the presence of impurities from sample to sample, arising from the fabrication process. The coherent fitting of Eq. 3 to all samples demonstrates that both spin-flip mechanisms are independent of the temperature [35] and of the thickness of Cu. The temperature dependence of the spin relaxation rate is thus given by the phonon scattering rate, whereas the temperature-independent part is given by the defect scattering rate, which increases with decreasing thickness and contributes to the spin relaxation rate at all temperatures. This is clearly observed in Fig. 4, where all the parallel linear curves are shifted up with decreasing Cu thickness.

## C. Discussion on the origin of the maximum in $\lambda_{Cu}$

Finally, we discuss the origin of the maximum in $\lambda_{Cu}$. From the previous discussion of the contributions to the spin relaxation in the framework of the Elliott-Yafet mechanism, the contribution of the phonons is the only responsible for the temperature dependence of the spin relaxation rate and therefore of $\lambda_{Cu}$. Accordingly, from the temperature dependence of the resistivity, an increase of $\lambda_{Cu}$ with decreasing temperature until saturation at low temperatures is expected. This is observed down to 30 K, where $\lambda_{Cu}$ starts to decrease with decreasing temperature (see Fig. 3(a)). This effect can only be explained by introducing a temperature dependence of the contribution of the defects, which include grain boundaries, surface and impurities.

Since we have already shown that the contribution of grain boundaries is temperature-independent, we could hypothesize that the observed temperature dependence arises from the surface contribution. This has been explicitly taken into account by Mihajlović *et al.* [5], who added an extra term of the form $a_s/\tau_e^s$ to Eq. 3. The temperature dependence arises from the fact that the surface scattering time $\tau_e^s$ is proposed to be inversely proportional to the one coming from the bulk. As a result, surface contribution to spin-flip scattering should dominate at low temperatures, when the mean free path becomes comparable to the dimensions of the NM nanowire, and the temperature at which the maximum of $\lambda_{NM}$ occurs should increase when decreasing the thickness of the NM nanowire. However, our results clearly show that the maximum of $\lambda_{Cu}$ always occurs at 30 K, regardless of the thickness of Cu (see Fig. 3(a)). Furthermore, the assumption that the surface scattering time is inversely proportional to the bulk one would necessarily imply an upturn in $\rho_{Cu}$ at low temperatures, which is not observed (Fig. 2).

A second option is that the temperature dependence comes from the impurities' contribution. In particular, magnetic impurities have not been considered in our previous analysis, since the Elliott-Yafet mechanism describes the spin-flip scattering in metals in the absence of such [4,34,35]. A temperature-dependent spin-flip probability coming from magnetic impurities is an option which is supported by recent studies made in Py/Cu [15] and Py/Ag [6] LSV devices. According to those works, the oxidation of the surface of the NM nanowire [15], or its capping with a MgO layer [6] induce the extinction of the maximum of the spin diffusion length. Such disappearance is attributed to the isolation from the NM nanowire of the magnetic impurities, which are most likely located at the surface due to the fabrication process [15]. Although the presence of magnetic impurities at the surface is unlikely in our case due to the fabrication process in two steps, magnetic impurities located at the bulk would yield the same effect. Further studies might be needed to quantify this contribution to spin-flip scattering.

## IV. CONCLUSIONS

In conclusion, we systematically measured both charge and spin transport in Cu nanowires as a function of temperature and thickness using lateral spin valves, in order to determine the different contributions to the spin relaxation. From a careful analysis based on the Elliott-Yafet mechanism, we found that the spin-flip probabilities coming from the phonons and the defects are both temperature and thickness independent. Whereas the temperature dependence of the spin relaxation is given by the phonon scattering as in bulk Cu, the temperature-independent part comes from defect scattering, which increases with decreasing thickness. Surprisingly, defect scattering in our Cu nanowires is clearly dominated by the grain boundaries rather than the surface. In addition, the maximum in the spin diffusion length of Cu observed at low temperatures, a puzzling feature reported before [5,6,14-16], cannot be explained by the Elliot-Yafet mechanism, suggesting a temperature-dependent spin-flip probability arising from magnetic impurities. Although additional spin relaxation contributions are unavoidable in confined nanostructures such as metallic nanowires, increasing the grain size or reducing the amount of magnetic impurities during the fabrication of spintronic nanodevices can be an efficient approach to overcome such limitations, leading to an improvement of the spin transport.

## ACKNOWLEDGEMENTS

This work is supported by the European Union 7th Framework Programme under the Marie Curie Actions (PIRG06-GA-2009-256470) and the European Research Council (Grant 257654-SPINTROS), by the Spanish Ministry of Science and Education under Project No. MAT2009-08494 and by the Basque Government under Project No. PI2011-1. E. V. and M. I. thank the Basque Government for a PhD fellowship (BFI-2010-163 and BFI-2011-106).

| *th (nm)* | $\rho_{Cu}$ *(μΩ cm)* | $\lambda_{Cu}$ *(nm)* | $\tau_e^{def}$ *(×10⁻³ ps)* | $a_{ph}$ *(×10⁻⁴)* | $a_{def}$ *(×10⁻⁴)* |
|---|---|---|---|---|---|
| 145 | 1.18 ± 0.01 | 1020 ± 6 | 35.8 ± 0.6 | 12.8 ± 0.8 | 8.2 ± 0.5 |
| 100 | 1.26 ± 0.01 | 1020 ± 46 | 33.5 ± 0.8 | 11.9 ± 0.9 | 7.0 ± 0.5 |
| 70 | 1.63 ± 0.01 | 863 ± 17 | 25.9 ± 0.8 | 11.2 ± 1.0 | 6.3 ± 0.6 |
| 60 | 1.98 ± 0.01 | 709 ± 120 | 21.3 ± 0.5 | 11.8 ± 0.5 | 6.8 ± 0.3 |
| 40 | 2.22 ± 0.01 | 501 ± 19 | 19.0 ± 0.2 | 9.5 ± 2.0 | 11.0 ± 0.3 |

Table 1: Summary of the most relevant data of the samples used in this work: thickness of the Cu channel, resistivity and spin diffusion length at 10 K, defect scattering time, and spin-flip probabilities from phonon and defect scattering.